\documentclass[twocolumn]{aastex631}

\shorttitle{Aerosols detection at microbar pressures}
\shortauthors{Estrela et al.}

\graphicspath{{./}{figures/}}

\begin{document}

\title{Detection of Aerosols at Microbar Pressures in an Exoplanet Atmosphere}

\author{Raissa Estrela}
\affil{Jet Propulsion Laboratory, California Institute of Technology, 4800 Oak Grove Drive, Pasadena, California 91109, USA}
\affil{Center for Radio Astronomy and Astrophysics Mackenzie (CRAAM), Mackenzie Presbyterian University, Rua da Consolacao, 896, Sao Paulo, Brazil}

\author{Mark R. Swain}
\affil{Jet Propulsion Laboratory, California Institute of Technology, 4800 Oak Grove Drive, Pasadena, California 91109, USA}
 
 
\author{Gael M. Roudier}
\affil{Jet Propulsion Laboratory, California Institute of Technology, 4800 Oak Grove Drive, Pasadena, California 91109, USA}

\author{Robert West}
\affil{Jet Propulsion Laboratory, California Institute of Technology, 4800 Oak Grove Drive, Pasadena, California 91109, USA}

\author{Elyar Sedaghati}
\affil{European Southern Observatory, Alonso de C\'ordova 3107, Santiago, Chile}
\affil{Instituto de Astrof\'isica de Andaluc\'ia (IAA-CSIC), Glorieta de la Astronom\'ia s/n, 18008 Granada, Spain}

\author{Adriana Valio}
\affil{Center for Radio Astronomy and Astrophysics Mackenzie (CRAAM), Mackenzie Presbyterian University, Rua da Consolacao, 896, Sao Paulo, Brazil}



\begin{abstract}


Formation of hazes at microbar pressures has been explored by theoretical models of exoplanet atmospheres to explain Rayleigh scattering and/or featureless transmission spectra, however observational evidence of aerosols in the low pressure formation environments has proved elusive. Here, we show direct evidence of aerosols existing at $\sim$1 microbar pressures in the atmosphere of the warm sub-Saturn WASP-69b using observations taken with Space Telescope Imaging Spectrograph (STIS) and Wide Field Camera 3 (WFC3) instruments on the Hubble Space Telescope. The transmission spectrum shows a wavelength-dependent slope induced by aerosol scattering that covers 11 scale heights of spectral modulation. Drawing on the extensive studies of haze in our Solar System, we model the transmission spectrum based on a scaled version of Jupiter's haze density profile to show that WASP-69b transmission spectrum can be produced by scattering from an approximately constant density of particles extending throughout the atmospheric column from 40 millibar to microbar pressures. These results are consistent with theoretical expectations based on microphysics of the aerosol particles that have suggested haze can exist at microbar pressures in exoplanet atmospheres. 
\end{abstract}





\section{Introduction} \label{sec:intro}

Aerosols have a critical role in establishing energy budgets, thermal structure, chemistry, and dynamics in planetary atmospheres \citep{Tomasko10, Muller14}. In the study of exoplanets, atmospheric aerosols have been suggested to explain the slope detected in optical and near-infrared (NIR) transmission spectra in some hot-Jupiters \citep{Pont08, Pont13, Mc14, Sing16} including flat atmospheric transmission spectra \citep{Gibson13,Knutson14b,Knutson14a,Kreidberg14}. Photochemical hazes can be produced from hydrocarbons \citep{Kempton17, Lavvas17} or sulfur-based particles \citep{Zah09}. Condensates \citep{Pinhas17, Morley13, Lecavelier08} composed of silicates or other exotic compounds are expected from thermochemical equilibrium calculations but often require vigorous diffusive or advective vertical transports to loft them to altitudes well above their cloud base \citep{Komacek19}.

Our own Solar System illustrates the diversity of roles and the physical origins of aerosols. On Jupiter and Saturn, hydrocarbon hazes formed by photochemistry extend through the stratosphere and upper troposphere, while clouds are present in deeper regions \citep{Atreya05, Nixon10}. A thick hydrocarbon haze layer on Saturn's moon Titan, present at $\sim$ millibar pressures \citep{Chai95}, cools the moon's surface by reflecting sunlight back into space \citep{Waite}. Our terrestrial neighbor Venus has photochemically produced hazes composed of H$_{2}$SO$_{4}$ in the upper layers of its atmosphere at altitudes of around 60 km (160 mbar) \citep{Titov18}. The multi-layered haze on Pluto detected by the New Horizons mission is believed to be cooling the planet's atmosphere and models suggest that it may exist at altitudes as high as 1000 km \citep[$\sim1\times10^{-11}$ bars;][]{Zhang17}. A thick haze layer similar to Titan's is believed to have been present on Earth during the Archean Eon, likely shielding life from ultraviolet (UV) light \citep{Trainer06, Arney18}, which could have allowed ammonia to build up, warming the Earth at a time that the young Sun was $\sim$20\% fainter than today \citep{Wolf10}. Shielding UV radiation is potentially important for the emergence of life on terrestrial planets as short wavelength UV radiation can be very harmful \citep{Estrela19, Estrela20}. 


Theoretical studies \citep{Lavvas17,Kawa18,Lavvas19,Kawa19,Adams19} using microphysics models that consider collisional growth, sedimentation, and transport by eddy diffusion of aerosol particles show that photochemical hazes based on hydrocarbons localizes the haze formation to be abundant at high altitudes (1\textendash10$\mu$bar). Alternatively, silicate condensates (MgSiO$_{3}$, MgSiO$_{4}$) can induce a strong optical scattering slope in hot giant exoplanets with equilibrium temperatures above 950 K \citep{Gao2020}, though it remains unclear if significant amounts of condensates can be lifted to $\sim \mu$bar levels.

Although theoretical models \citep[e.g.][]{Lavvas19} make clear predictions of photochemical aerosol formation at $\sim\mu$bar pressures, obtaining direct observational evidence of aerosols in exoplanet atmospheres at these pressures is challenging due to the pressure-composition ambiguity affecting the choice of effective radius \citep{Griffith14, Mieux17}. Here, we report on observations that probe these pressure regions through the use of broad spectral coverage and the presence of a NIR water feature to reduce the pressure-composition ambiguity. Vertical profiles of photochemical hazes in our Solar System are more diverse than what has been used typically in the exoplanet community; motivated by this awareness, we interpret the transmission spectrum of WASP-69b using a scaled version of the formalism developed by \cite{Zhang15} in their study of Jovian aerosols. 


\section{Methods}

Our study reports on observations of WASP-69b which orbits a K5 star with an orbital period of 3.868 days \citep{And14}. WASP-69b has a large atmospheric scale height of $\sim$650 km and transits a bright host star (V $\sim$ 10), making it a favorable target for atmospheric detection. We performed a uniform analysis of the composite spectrum of WASP-69b using archival data obtained with the Space Telescope Imaging Spectrograph (STIS) and Wide Field Camera 3 (WFC3) instruments on-board the Hubble Space Telescope (HST). To interpret the uniformly processed spectrum, we use a Bayesian atmospheric retrieval that includes thermochemical equilibrium and disequilibrium models. We used observations taken with HST/STIS in both G430L (0.3\textendash0.5 $\mu$m) and G750L (0.5\textendash1.0 $\mu$m) visible grisms and G141 infrared grism transit observations of WASP-69b available in the Mikulski Archive for Space Telescopes (MAST) archive (Proposal ID 14767, PI: David Sing). We use our data reduction and analysis pipeline, EXCALIBUR \citep[EXoplanet CALIbration and Bayesian Unified Retrieval;][]{Swain21}, which performs an automatic and fully algorithmic data reduction of a data input catalog observed with HST/STIS G430L and G750L grism available from MAST. EXCALIBUR is also implemented with the CERBERUS software \citep{Roudier21} that performs atmospheric retrieval by implementing a Bayesian scheme for efficient evaluation of thermal equilibrium versus disequilibrium model classes.

EXCALIBUR tasks can be divided into three major steps: spectral extraction, light curve modeling, and spectral retrieval. We focus here on the STIS-related aspects of the EXCALIBUR data reduction; data reduction of WFC3 spatial scan observations is previously described in \cite{Swain21,Roudier21}. Prior to processing by EXCALIBUR, the raw STIS data are pre-processed for bias, dark current and flat field corrections using the latest version of the Space Telescope Science Institute (STScI) CALSTIS\footnote{https://stistools.readthedocs.io/en/latest/} pipeline provided by MAST. The STScI CALSTIS data products are the starting point for the spectrum extraction step.

\subsection{Spectrum Extraction}

Using the 2D images from consecutive reads acquired during the detector integration time, we obtain 1D spectra by summing each column from the image. We use the median exposure length of each visit as a reference to ignore the exposures with very low counts. Before obtaining the 1D spectra, we remove cosmic rays in the 2D image and a significant fringing effect that affects the G750L grism \citep{Nikolov14, Loth18}. We perform fringing correction in G750L using a library of contemporaneous flat fringes that were taken during the observations of the target. This process requires removal of the lamp function from the contemporaneous flat fringes by applying a polynomial fit of order 11 to each row of the observation following the method of \cite{Plait97}. We normalize the stellar spectrum by dividing it by its continuum to show the relative amplitude of the fringes and finally we divide the spectra by the flat field only for the region mostly affected by the fringes ($>$ 0.8 $\rm \mu$m). 

STIS data contain a large number of cosmic ray events that we remove using the following two steps: 1) flag the 2D image by subtracting the image from its median filter, points above 3$\rm \sigma$ are replaced by the median filter correspondent value; and 2) apply a sigma clipping technique in the 1D spectra to reject residual cosmic rays.

\begin{figure*}[!ht]
\centering
\includegraphics[width=0.8\textwidth]{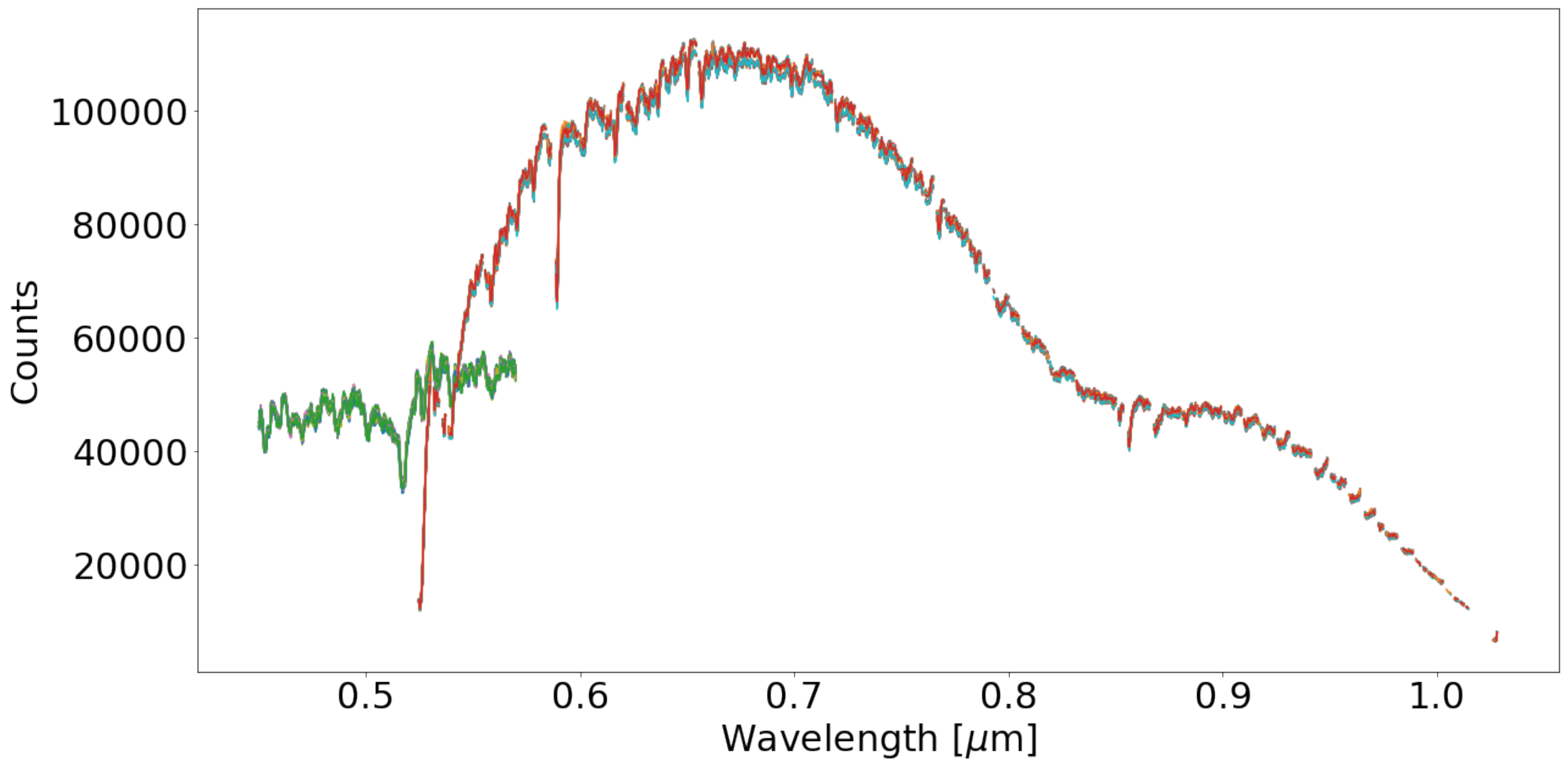}
\caption{Wavelength calibrated HST/STIS spectra with EXCALIBUR for grisms G430L and G750L.}
\label{fig:cal}
\end{figure*}

\begin{figure}[!ht]
\centering
\includegraphics[width=0.5\textwidth]{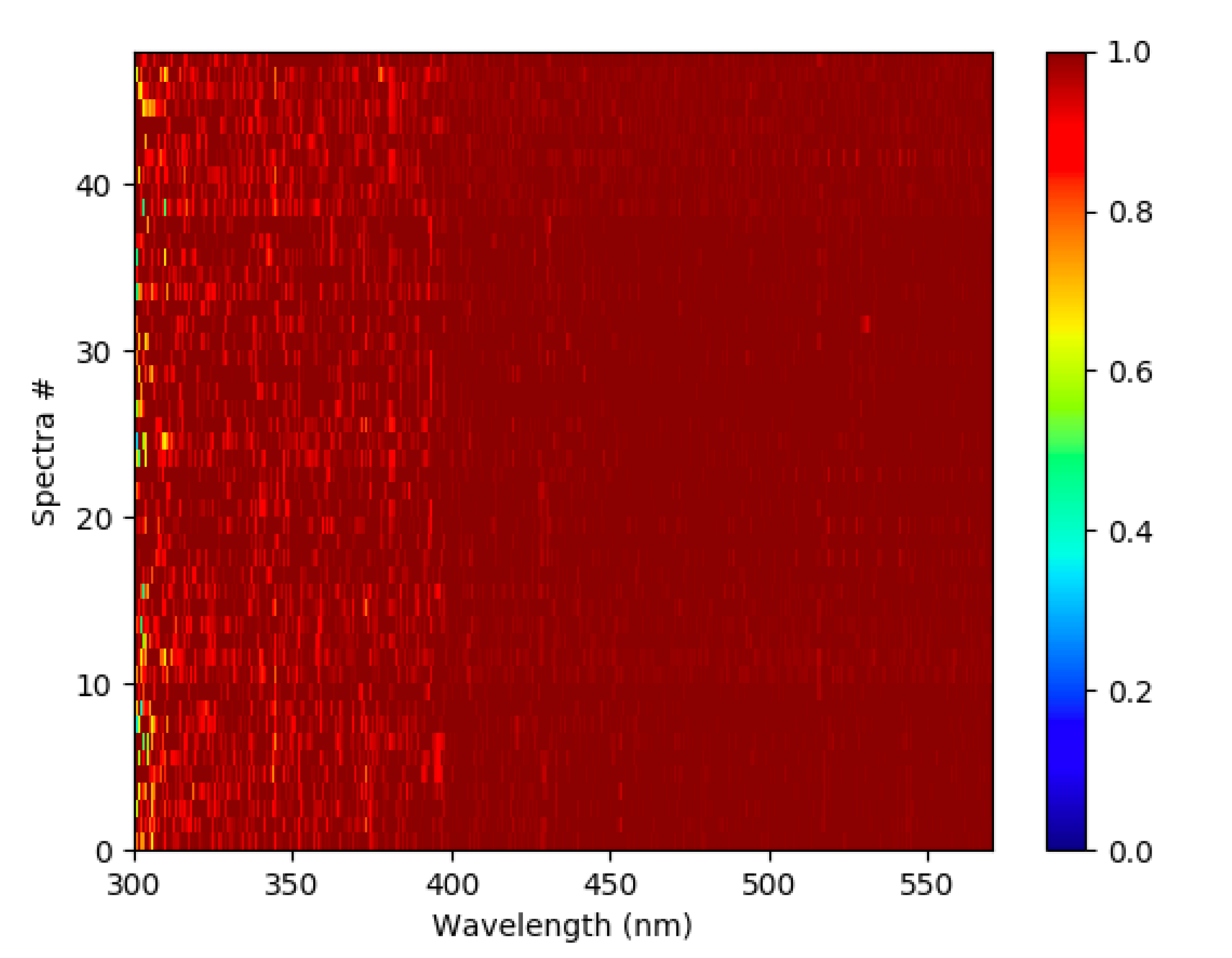}
\caption{A total of 48 frames of the 1D spectra flattenned and stacked in an image to highlight the noise that dominates the region between 0.3-0.4$\mu$m in HST/G430L.}
\label{fig:noise}
\end{figure}

Because of the large difference in the effective shape of the spectra for the G750L and G430L filters, the wavelength calibration is performed in two different ways depending on the grism. For G750L, as a model for the stellar spectrum, we use the logarithmic of the throughput of the filter times a blackbody function of the star, because using the logarithm is less sensitive to model discrepancies due to absorption lines. We shift and scale the spectrum from the individual pixel-based rows to match the stellar spectrum model so that the steep gradients at the edges of the spectrum are aligned. In the case of G430L, we adopt a different calibration process for two reasons: 1) the strong absorption feature due to MgH at $\sim$545 nm overlaps strong broad stellar absorption bands \citep{Weck03} that are present in the spectrum and are not reproduced by the blackbody function of the star; and 2) the throughput of the filter has a lower efficiency that does not reproduce well the shape at the red end of the observed spectra. Therefore, we adopted the more robust PHOENIX stellar spectra model, which comes in a grid with a vast range of $T_{eff}$, $\log g$, and $[Fe/H]$ \citep{Husser13}. For each star, we chose the PHOENIX model that has the closest physical parameters. Then, we multiply the model with the throughput of the instrument to reproduce the observed spectrum. Finally, for G430L, we repeat the procedure from G750L by shifting and scaling each observed spectrum to match the spectrum model. The final calibrated spectra in both filters are shown in Fig.~\ref{fig:cal}. Also, in G430L we exclude the data between 300\textendash400 nm because this region is dominated by noise, as evident in the flattened and stacked 1D spectra (of which there are 48), presented in Fig. \ref{fig:noise}. 

\subsubsection{White Light Curves}
Once the extracted spectra are wavelength calibrated and trimmed in the spectral dimension to exclude high-noise regions, the spectra can be assembled into light curves for each visit. To obtain the transmission spectra, EXCALIBUR performs a two-step process. First, we obtain the broadband light curve using all available visits for each instrument/observing mode, which in the case of these observations includes STIS-G430, STIS-G750, and WFC3-G141. The broadband light curves for each observing mode are created by averaging the spectra in the wavelength axis. Next, an outlier rejection algorithm flags wavelength channels of the spectrum that are not stable relative to three times the expected shot noise level in the out-of-transit portion of the signal. We model the broadband light curve model following the \cite{Roudier21} approach, and derive estimates for the transit depth (R$_{p}$/R$_{s}$)$^{2}$, including the effects of the instrument model and the stellar limb darkening. 

The instrument model takes into account the systematics trends in the calibrated spectra  \citep[for details see][]{Swain21}. The main systematic trends are due to the thermal breathing of HST which warms up the spacecraft during its orbital day and cools it down during the night. This compromises the HST focus and mainly affects the first orbit of each HST visit \citep{Hasan93,Sing11}. Therefore, in line with previous studies \citep{Nikolov15}, we reject the first orbit of each visit. The instrument systematic model has independent linear components for the visit and orbit slope augmented by an orbit exponential ramp \citep{Kreidberg14}. 

\begin{figure}[!ht]
\centering
\includegraphics[width=0.5\textwidth]{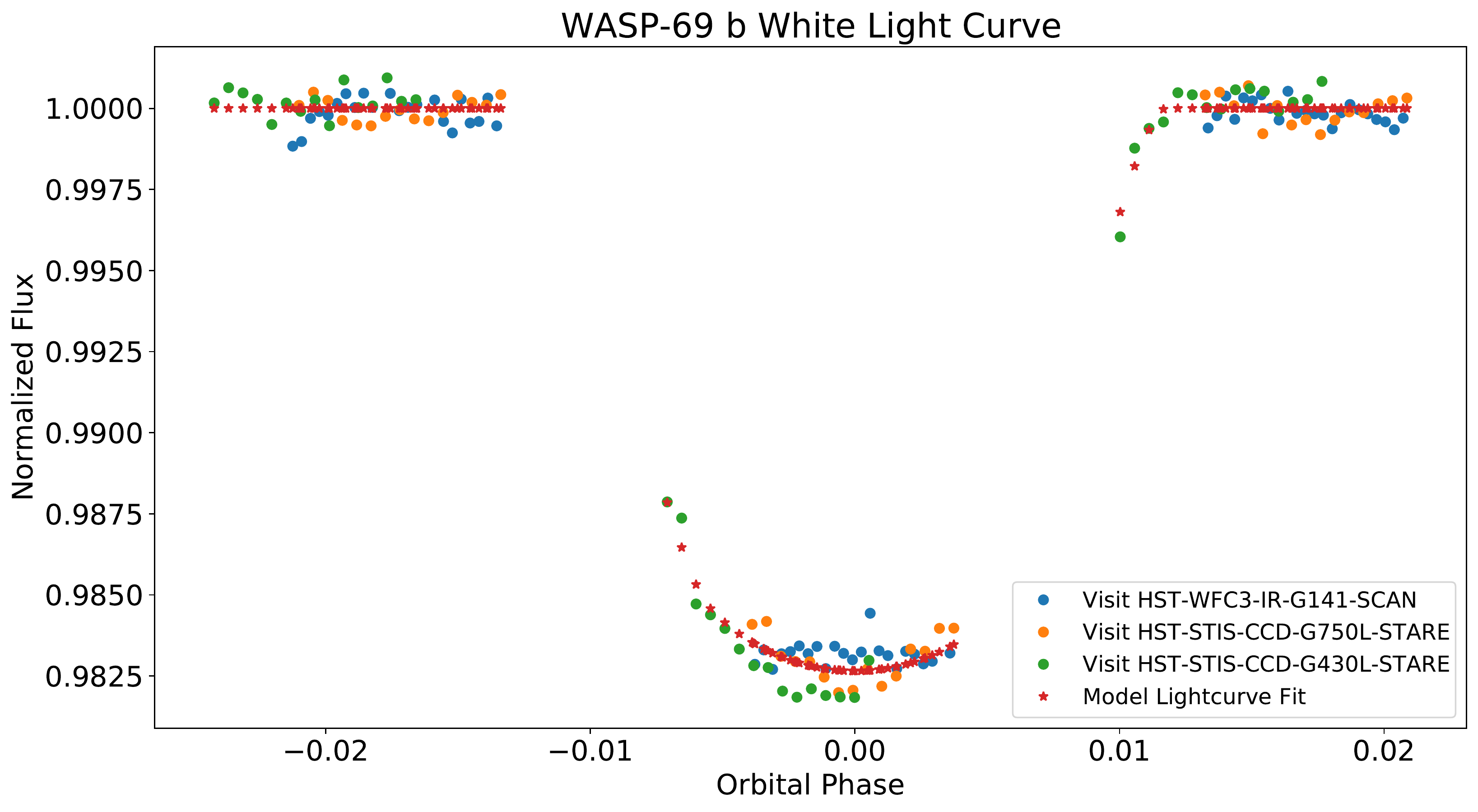}
\caption{Broadband white light curve of WASP-69b for HST/STIS G430L (green), HST/STIS G750L (orange), and HST/WFC3 G141 (blue), and the retrieved transit model is indicated in red. }
\label{fig:broadband}
\end{figure}

The transit model is described in terms of the exoplanet's orbital eccentricity $e$, impact parameter $b$, semi-major axis $a$, period $P$, inclination $i$, time of mid-transit per visit $T_{0}$ and the relative planetary radius $R_{p}/R_{s}$. The best fitting transit parameters for the white light curve fit are
given in Table \ref{tab:system}. We solve for the system parameters using the multi-filter light curve, which makes our solution self-consistent for G430L+G750L+G141. Our pipeline also incorporates a four parameter non-linear limb-darkening model from \cite{Claret00} that accurately describes the intensity distribution with limb angle. To retrieve the orbital solution, the instrument model and the limb darkening coefficients, the broadband transit light curve model is used in conjunction with a Metropolis-Hastings sampler from PyMC3 package \citep{Salvatier16} and uses orbital parameters from the NASA Exoplanet Science Institute (NExScI) database as \textit{a priori} information. Fig. \ref{fig:broadband} shows the broadband light curves obtained for all the HST filters used in this study.

\subsubsection{Spectral Light Curves}
\begin{figure}[!ht] 
\centering
\includegraphics[width=0.45\textwidth]{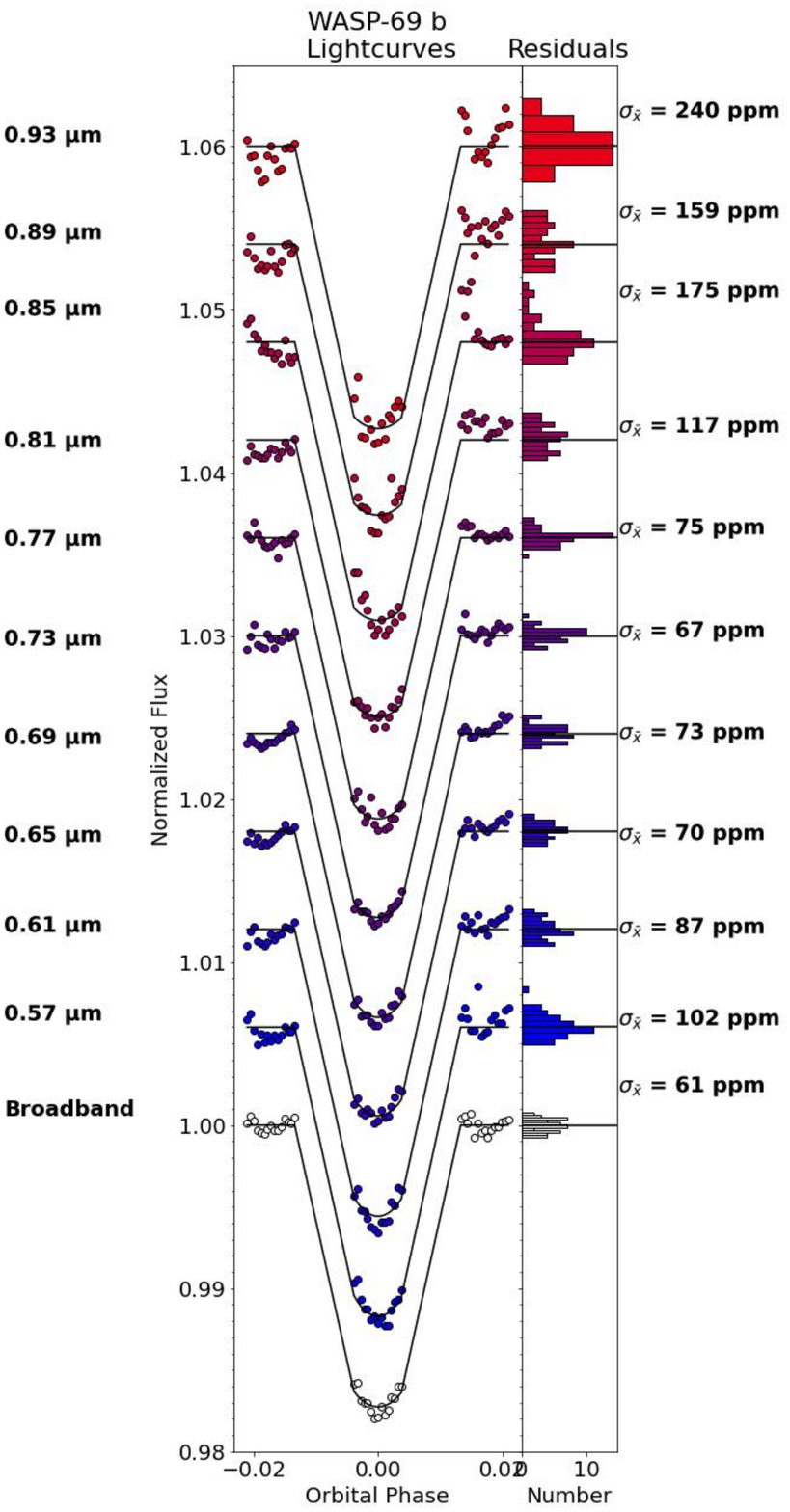}
\caption{White light and spectral light curves associated with the calibration and modeling of the Hubble STIS/G750L observations. The stronger impact of limb darkening at shorter wavelengths is evident in the STIS in transit light curve segment.}
\label{fig:LCs}
\end{figure}

For the second step of our process, we create a wavelength template with 105 and 121 spectral points for G750L and G430L, respectively, and sum the spectral points within each wavelength bin to obtain the spectral light curves. In constructing the spectral point light curves, we average 3 spectral pixels for G430L and 7 spectral pixels for G750. The instrument Line Spread Function \citep{STIShandbook} full width half max (FWHM) is approximately 1.31 pixels for G430L and approximately 1.46 pixels for G750L. Each of the 100 spectral point light curves are independent spectral samples because the number of pixels averaged corresponds to a wavelength interval that is larger than the instrument Line Spread Function. For the spectral light curves, we follow a similar approach to modeling the broadband light curve except that we use the retrieved orbital solution values of all parameters that are not wavelength dependent (impact parameter and time of mid transit) and retrieve the transit depth (R$_{p}$/R$_{s}$)$^{2}$, limb darkening coefficients, and the instrument model \citep[see description in][for more details]{Swain21}. The spectral light curves and their corresponding models are shown in Fig. \ref{fig:LCs}.  
 
\subsection{Atmospheric Retrieval: CERBERUS}

CERBERUS is a spherical shell radiative transfer modeling code coupled with parameter retrieval using a Markov Chain Monte Carlo (MCMC) approach and model comparison capabilities. Details of CERBERUS capabilities are described in \cite{Swain21} and \cite{Roudier21}. Two families of atmospheric models are considered here. One is the thermal equilibrium chemistry (TEC), in which the forward model is using a standard parametrization of the exoplanet atmospheric atomic reservoir (X/H, C/O and N/O), and that computes the vertical mixing ratio (VMR) of molecular species according to thermal equilibrium chemistry. The other model is disequilibrium chemistry that includes a limited number of dominant absorbers (CH$_{4}$, C$_{2}$H$_{2}$, HCN, CO$_{2}$), with reaction products mainly due to photochemistry and transport.

\begin{figure}[!ht]
\centering
\includegraphics[width=0.45\textwidth]{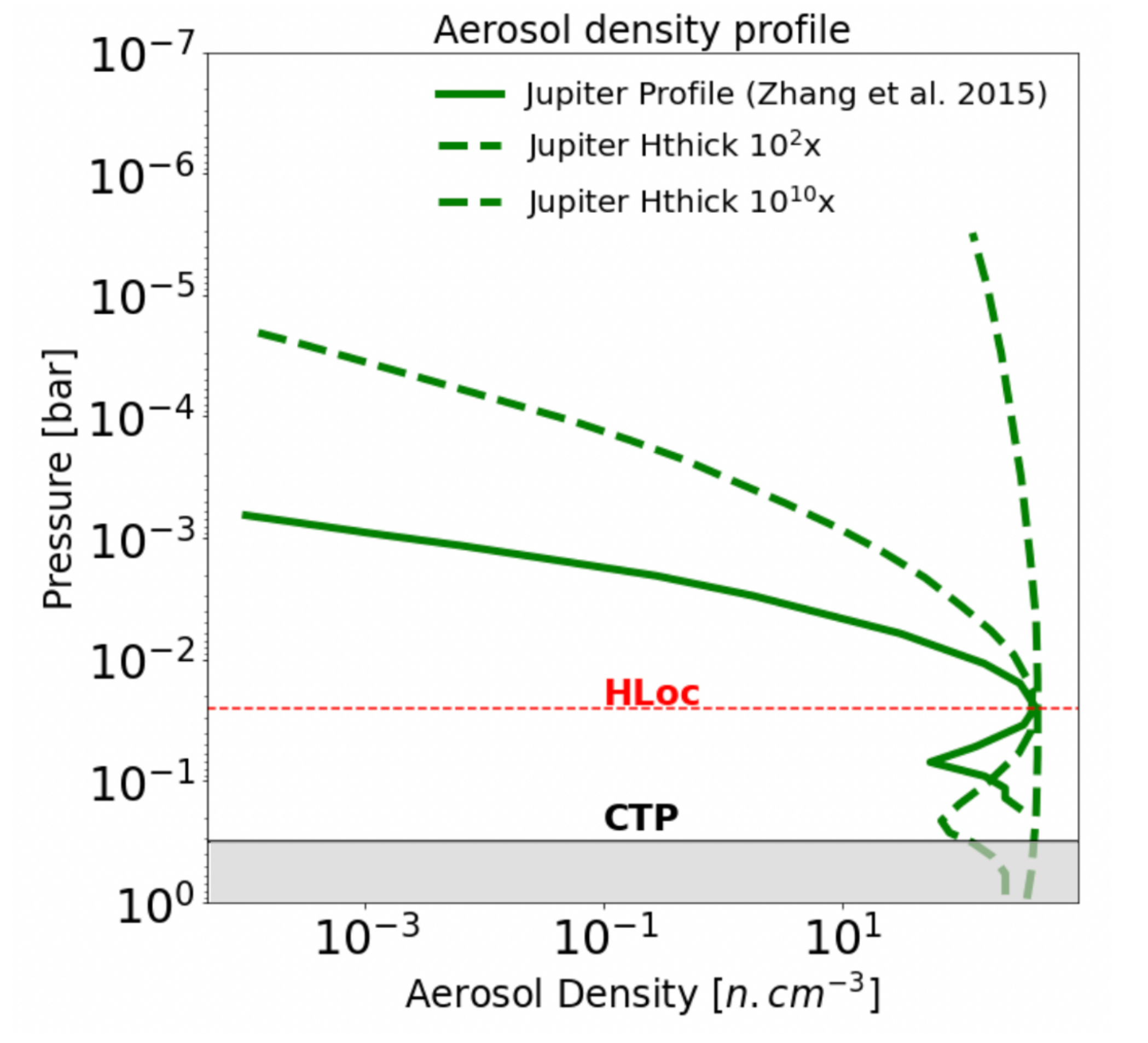}
\includegraphics[width=0.45\textwidth]{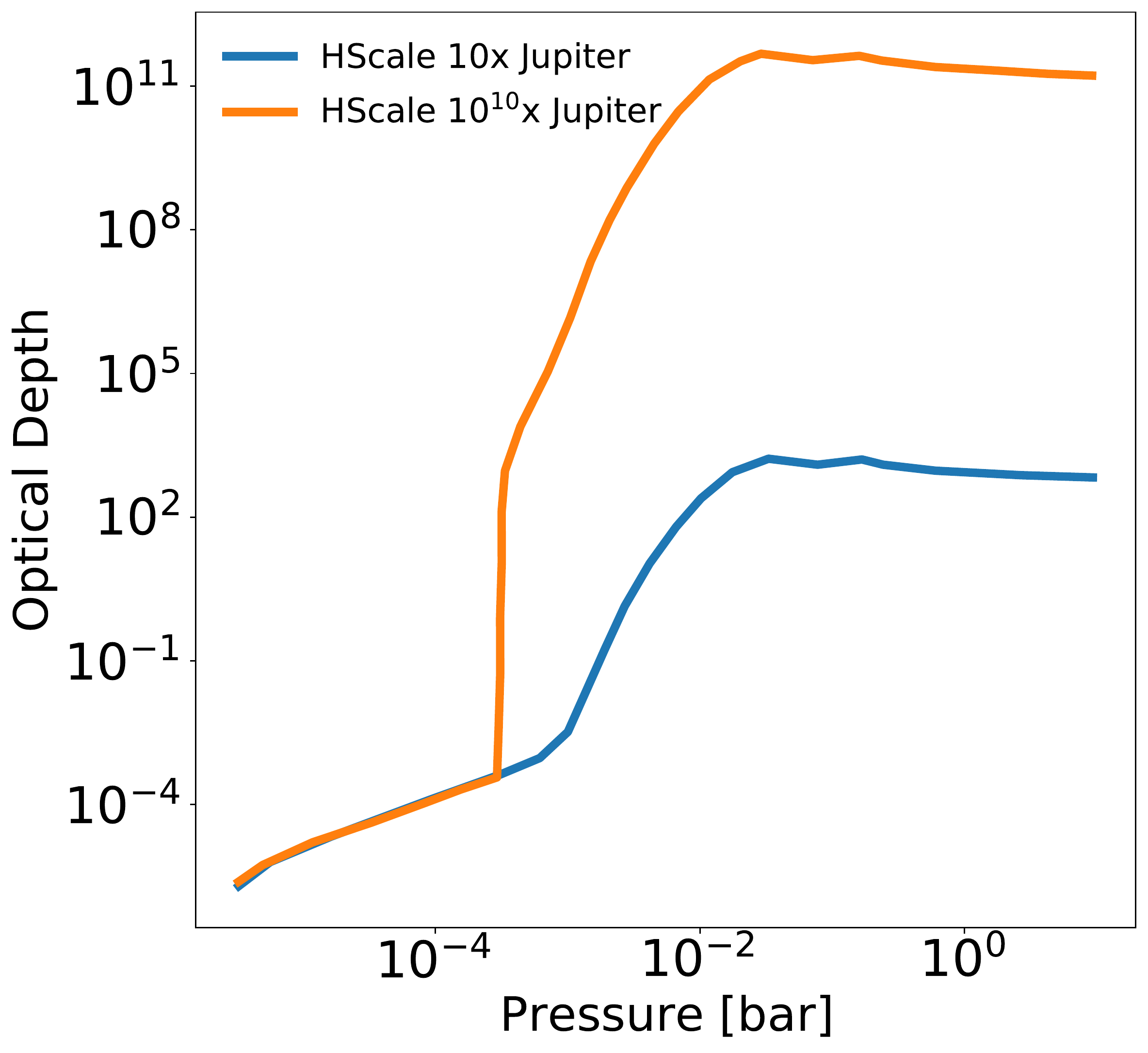}
\caption{Illustration of the impact of varying CERBERUS haze parameters. \textit{Left:} changes in the aerosol density profile by modifying CERBERUS parameter Hthick, which is a multiplicative factor of the width of the Jupiter haze density profile. The location of the peak density of the Jupiter haze profile (HLoc) is indicated by the red dashed line and the location of the cloud top pressure (CTP) is indicated by a solid line. All incidental radiation below the CTP is blocked and is represented by the grey area. \textit{Right:} Variations in the optical depth scaling parameter HScale.}
\label{fig:cerberuspar}
\end{figure}

The altitude range considered for this work spans from 10 bars, taken as the solid radius, R$_{0}$ \citep{Griffith14}, up to 20 scale heights (Hs) above, evenly log-divided into 100 discrete isothermal pressure layers in hydrostatic equilibrium. Gaseous sources are modeled based on opacities from EXOMol (H$_{2}$O, CH$_{4}$, HCN, H$_{2}$CO, CO, CO$_{2}$) \citep{Ten16}, HITEMP (C$_{2}$H$_{2}$, N$_{2}$, N$_{2}$O, O$_{2}$, O$_{3}$, TiO) \citep{Roth10}, and on collision induced opacities from HITRAN (H$_{2}$ – H$_{2}$, H$_{2}$ – He, H$_{2}$ – H, He – H) \citep{Richard12}, and we assume a constant temperature profile and a constant vertical mixing ratio (VMR) for gaseous species.

Rayleigh scattering from molecular hydrogen, H$_{2}$, is computed by assuming a constant vertical mixing ratio, taken as solar throughout the considered atmosphere. The equivalent wavelength dependency of the cross section follows:

\begin{equation}
    S_{H_{2}} = S_{0} \left( \frac{\lambda}{\lambda_{0}} \right)^{-4}
\end{equation}

\noindent assuming $S_{0} = 2.52 \times 10^{-28}$ $ \times 10^{-4}$  m$^{2}$ mol$^{-1}$ and $\lambda_{0}$ = 750 $\times$ 10$^{-3}$ $\mu$m \citep{Naus0}. We account for scattering through aerosols by parametrizing the haze density profile using an aggregate of Jupiter's haze profiles \citep{Zhang15} constructed by taking the median of the haze latitude number density. The following free parameters characterize the haze properties and are retrieved by CERBERUS:

\begin{itemize}
\item CTP: bottom opaque cloud top pressure in log (bars)
\item HScale: multiplicative scale factor (log10) of the haze optical depth (common to all pressure layers). In CERBERUS, the optical depth per length element $\tau$/dl [m$^{-1}$] is given by:
\begin{equation}
   \rm \frac{\tau_{cerb}}{dl} = \rm HScale \times \rho_{cerb} \times \sigma_{cerb}
\end{equation}
where $\rho_{cerb}$ and $\sigma_{cerb}$ are the haze density profile and cross section in CERBERUS, respectively.
\item Hloc: the maximum density location of the haze cloud in log pressure (bars)
\item Hthick: multiplicative factor (log10) of the width of the density haze profile 
\end{itemize}

The aerosol density profile has a log-pressure maximum density location (HLoc) that is allowed to linearly translate vertically in the atmosphere. A multiplicative factor (HThick) is introduced to compress or stretch the vertical extent of the initial profile and is expressed in H$_{s}$. Another multiplicative factor (HScale) is used to scale the haze optical depth for each pressure layer. This is a product of the cross section, density profile, and optical path length, assuming the Jupiter averaged haze cross section and density profile as a reference \citep{West04}. We account for clouds by taking them as surfaces that block ($\tau_{clouded}$ = 0) all incident radiation below the CTP. Figure \ref{fig:cerberuspar} illustrates the impact of varying the CERBERUS haze parameters on the haze profile when they are varied away from the baseline Jupiter case.

\section{Results} \label{sec:results}

\begin{figure*}[!ht]
\centering
\includegraphics[width=0.7\textwidth]{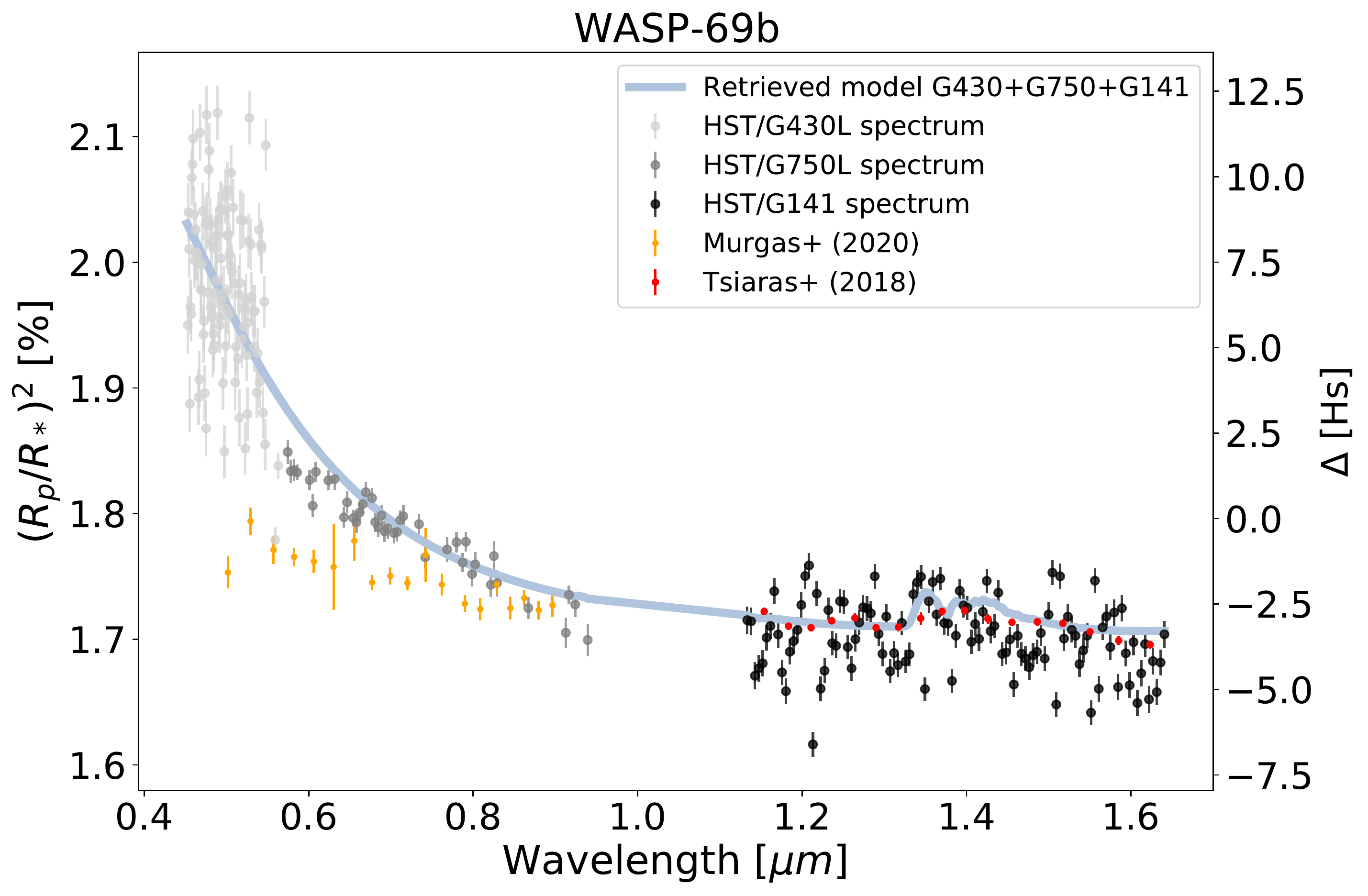}
\caption{Composite HST/STIS G430L, G750L, and HST/WFC3 G141 transmission spectra obtained with our uniform processing pipeline EXCALIBUR and atmospheric model retrieved by CERBERUS (blue) indicating a very steep slope associated with the presence of haze.}
\label{fig:mergedspec}
\end{figure*}

We find that the transmission spectrum of WASP-69b is dominated by a wavelength-dependent slope consistent with a Rayleigh scattering curve that is very pronounced in the two visible HST/STIS G430L and G750L filters from 0.4 to 1.0 $\mu$m. This curve extends to the NIR, covering approximately 11 Hs of spectral aerosol-induced modulation, as shown in Fig. \ref{fig:mergedspec}. To our knowledge, these data represent one of the greater continuum opacity samplings of the atmosphere of an exoplanet and underscore the variety of conditions that exist in the exoplanet population. The aerosol scattering slope changes by a factor of $\sim$ 5 from the visible to NIR wavelengths and is explained by a uniform density profile that is present from $\sim$ 40 millibar pressures to $\sim$microbar. This detection was made using observations taken with HST/STIS in both G430L (0.4 \textendash0.5 $\mu$m) and G750L (0.5\textendash1.0 $\mu$m) visible grisms and HST/WFC3 G141 (1.08\textendash1.68 $\mu$m) infrared grism from MAST. 

\begin{figure*}[!ht]
\centering
\includegraphics[width=0.65\textwidth]{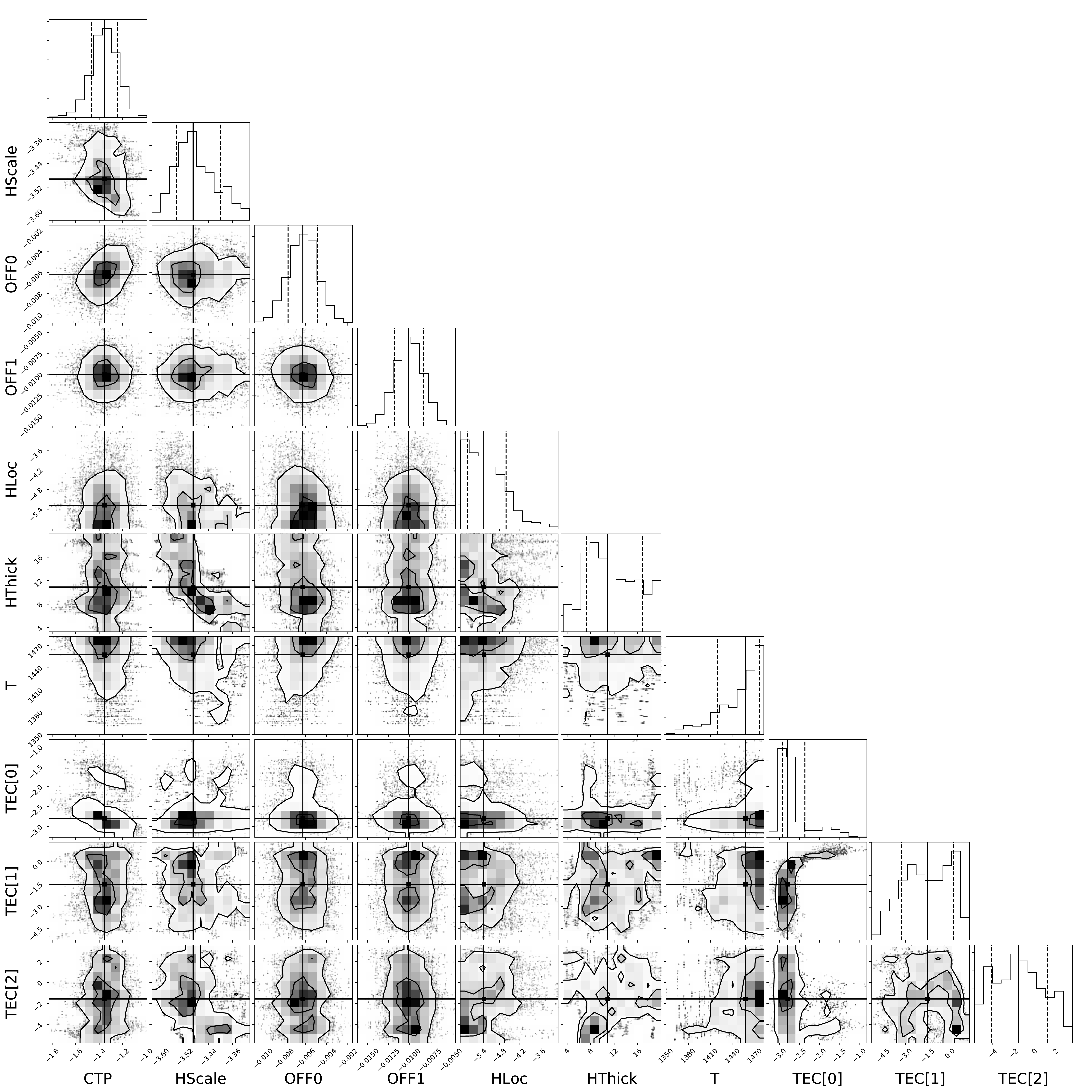}
\caption{Correlation plots and marginalized posterior distributions for retrieved atmospheric model parameters. HScale, HLoc, and HThick are all haze parameters corresponding,
respectively, to the multiplicative scale factor (log10) of Jupiter's haze-density profile, the maximum density location in log pressure (bars), and a multiplicative factor (log10) of the width of the Jupiter haze. T corresponds to the retrieved atmospheric temperature (K), and TEC[0], TEC[1], and TEC[2] label values correspond to the X/H, C/O, and N/O (log10) of the scaling factor applied to the reference solar value (0 $\equiv$ solar, 1 $\equiv$ 10 $\times$ solar). The vertical lines correspond to the average (solid) and $\pm$ 1$\sigma$ confidence interval values (dashed).}
\label{fig:corr_plots}
\end{figure*}

\begin{figure}[!ht]
\centering
\includegraphics[width=0.48\textwidth]{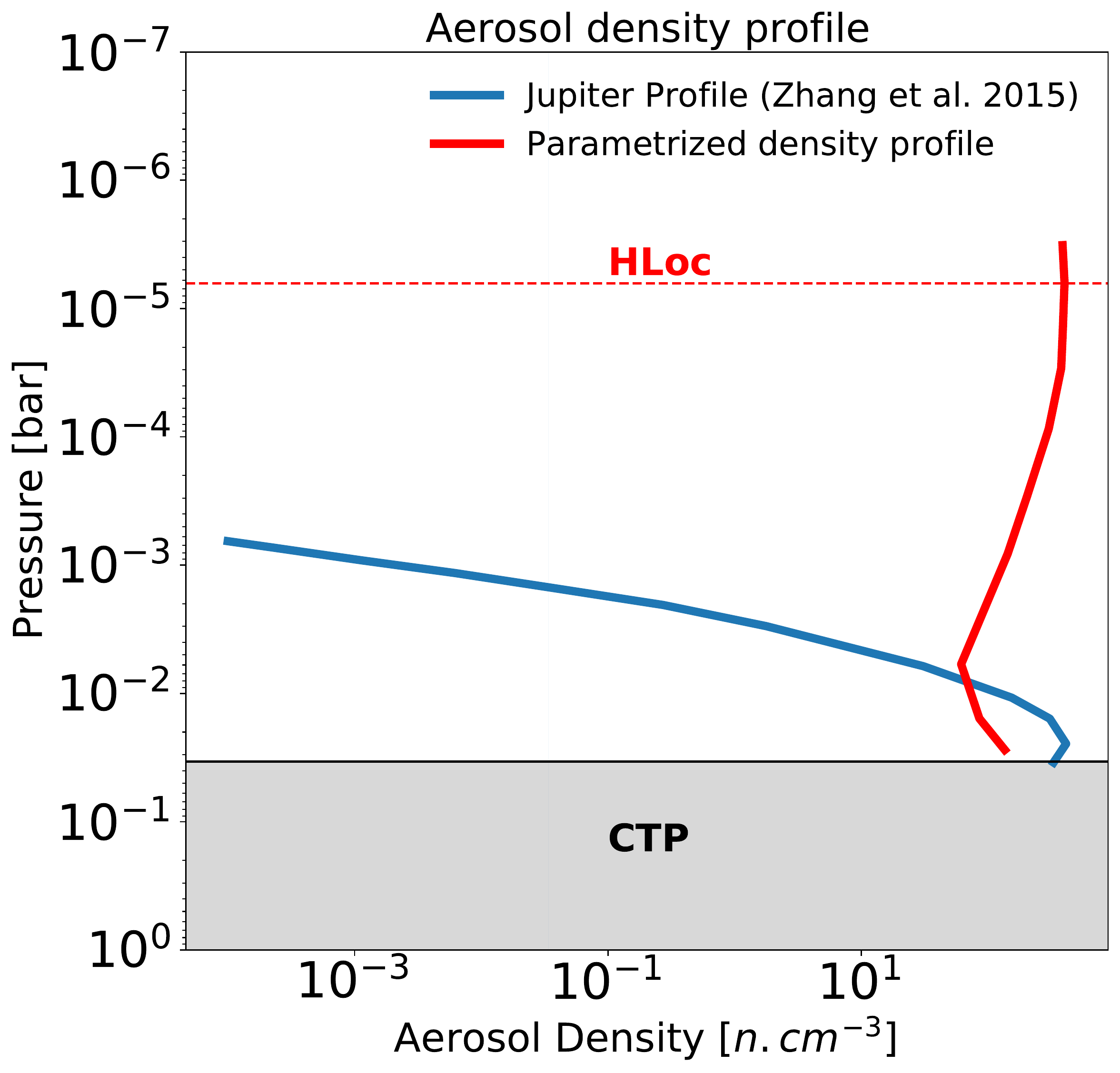}
\caption{Aerosol density profile retrieved by CERBERUS for WASP-69b. The location of the peak density of the aerosols (HLoc) is indicated by a dashed line. The bottom cloud (CTP) is indicated by a solid black line and anything below the CTP (within the grey region) would be muted. The range of pressure shown in this figure corresponds to the range of pressures modeled in our analysis.}
\label{fig:aerosol}
\end{figure}

A molecular feature is also observed at 1.4 $\rm \mu$m, which is suggestive of water presence that could be muted by the presence of aerosols. The water detection and the NIR spectra are in agreement with \cite{Tsiaras18} who analysed the same NIR dataset, although their result is based on averaging several pixel-based spectral channels together. To infer the atmospheric content in the observations, we used the CERBERUS atmospheric retrieval code, modeling the spectrum using a thermal equilibrium chemistry by using a standard parametrization of X/H, C/O, and N/O. Here we present the results obtained with the TEC model, although both TEC and disequilibrium models were tried. The TEC model provided a better agreement with the data with a delta BIC of 6, which provides strong evidence against the disequilibrium model. The retrieval parameter correlation plots (Figure \ref{fig:corr_plots}) show constraints for most parameters retrieved, but it does not give a well constrained C/O and N/O. The modeling is performed assuming an isothermal atmosphere and the retrieval results in an atmospheric model composed of a uniform haze layer that extends from CTP at 0.03 bars to lower pressures in the upper atmosphere ($\sim$ microbars), with a peak density at 6.3 $\rm \mu$bars (see Fig. \ref{fig:aerosol}). Any other molecular feature or aerosols present below the cloud would be muted. Although we use an isothermal atmospheric profile, we checked the dependence of the cross section with temperature for all the opacities sources (H$_{2}$O, NH$_{3}$, CO, CO$_{2}$, CH$_{4}$ and N$_{2}$) included in our TEC (Thermal Equilibrium Chemistry) model and find that none of them can add significant opacity at short wavelengths (400-500 nm), regardless of the temperature (for a range of temperatures between 300 and 2000 K). This finding emphasizes that haze is required to explain our observations as there is no other source of sufficient opacity we can identify that can explain the blue portion of the spectrum.In addition, departures from an isothermal profile still require the presence of haze, as shown in \cite{Lavvas21}, as haze can induce temperature inversion by heating the upper atmosphere and cooling the lower atmosphere.

We find sub-solar metallicity [X/H] values (0.001x solar) with our retrieval and further investigation on its implication for the formation of uniform hazes distribution is necessary. A collection of other gaseous exoplanets also show sub-solar metallicities. In particular, when compared to planets with similar mass to WASP-69b, such as HAT-P-12b and WASP-31b, that has metallicities values of 0.4$_{-0.38}^{+2.60}$ $\times$solar and 0.2$_{-0.03}^{+0.90}$ $\times$solar \citep{Mac19}. The haze parameters (HScale, HLoc, and HThick) in the retrieval are the ones that control the wavelength dependent slope seen in the spectrum, and the curvature of the slope is controlled by the multiplicative factor of Jupiter's optical depth (HScale). 

Two possible sources of aerosol opacity are hydrocarbon photochemical haze and silicate aerosols. A previous study \citep{Gao2020} found that for planets with equilibrium temperature (Teq) $>$ 950 K, like WASP-69b,  silicate aerosols are the main opacity; they also found that planets with Teq $<$ 950K, normally have an increase in methane abundance leading to hydrocarbon aerosols as the dominant source of opacity. However, the temperature retrieved ($\sim$1450 K) is much higher than the equilibrium temperature of the planet (963 K), which is qualitatively consistent with recent studies by \cite{Lavvas21} that suggest that photochemical hazes can induce heating of the upper atmosphere. On the other hand, our retrieved C/O suggests that significant C/O enhancement above solar values is inconsistent with the data and consequently a composition that is more consistent with condensates. The nature of these aerosols is further explored in Section \ref{sec:disc}.



\subsection{Rayleigh Scattering by Molecular Hydrogen}
\begin{figure}[!ht]
\centering
\includegraphics[width=0.5\textwidth]{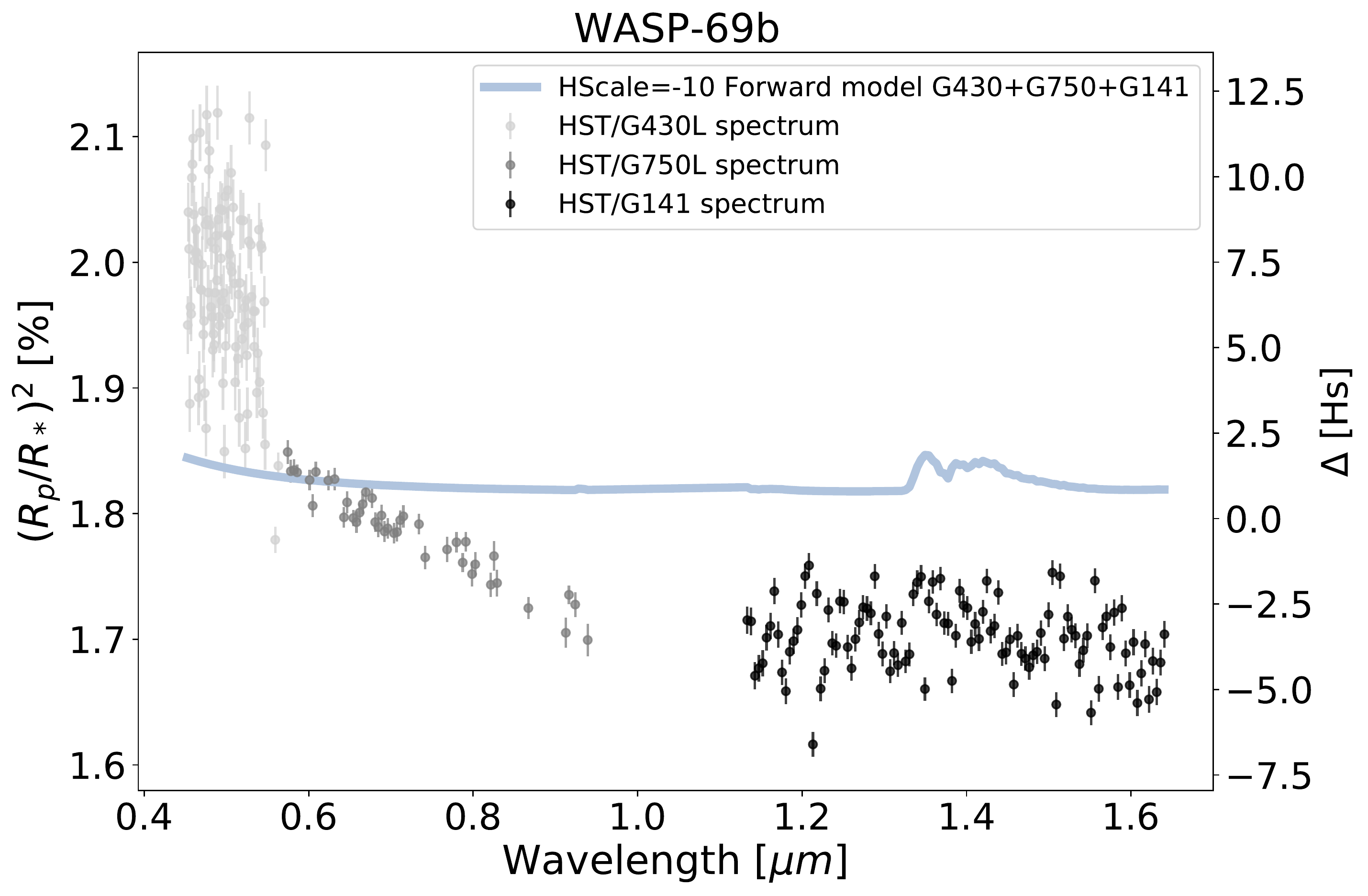}
\caption{Composite transmission spectra (grey) and forward model (blue) forced to have a scaling factor of the optical depth (Hscale) value that is outside of the window found by the atmospheric retrieval. This simulates H$_{2}$ scattering and shows that the measured spectrum requires an aerosol.}
\label{fig:flatmodel}
\end{figure}

Rayleigh scattering can also be caused by molecular hydrogen \citep{Lecavelier08}, however this is not the case in our present work. We confirm that the slope in the transmission spectrum is related to scattering due to haze because the CERBERUS retrieval shows that there is a very specific window in the haze optical depth scaling factor (HScale) that reproduce the slope, between -3.49 and -3.54 (see values of HScale in Table \ref{tab:retrieval} and triangle plots in Fig. \ref{fig:corr_plots}). If we use an optical depth scaling factor (HScale) value that is outside of the retrieval range (HScale = -10), we enter in the H$_{2}$ scattering domain and we lose the slope obtaining a flat model (see Fig. \ref{fig:flatmodel}).  We calculated the BIC for both models and find that the haze-free flat model has a BIC of -91, while the retrieved model has a lower BIC of -110. These values give a delta BIC > 10 ,which implies in very strong evidence against the haze-free model.


\subsection{Consistency with Ground-based Observations}
\begin{figure}[!ht]
\centering
\includegraphics[width=0.50\textwidth]{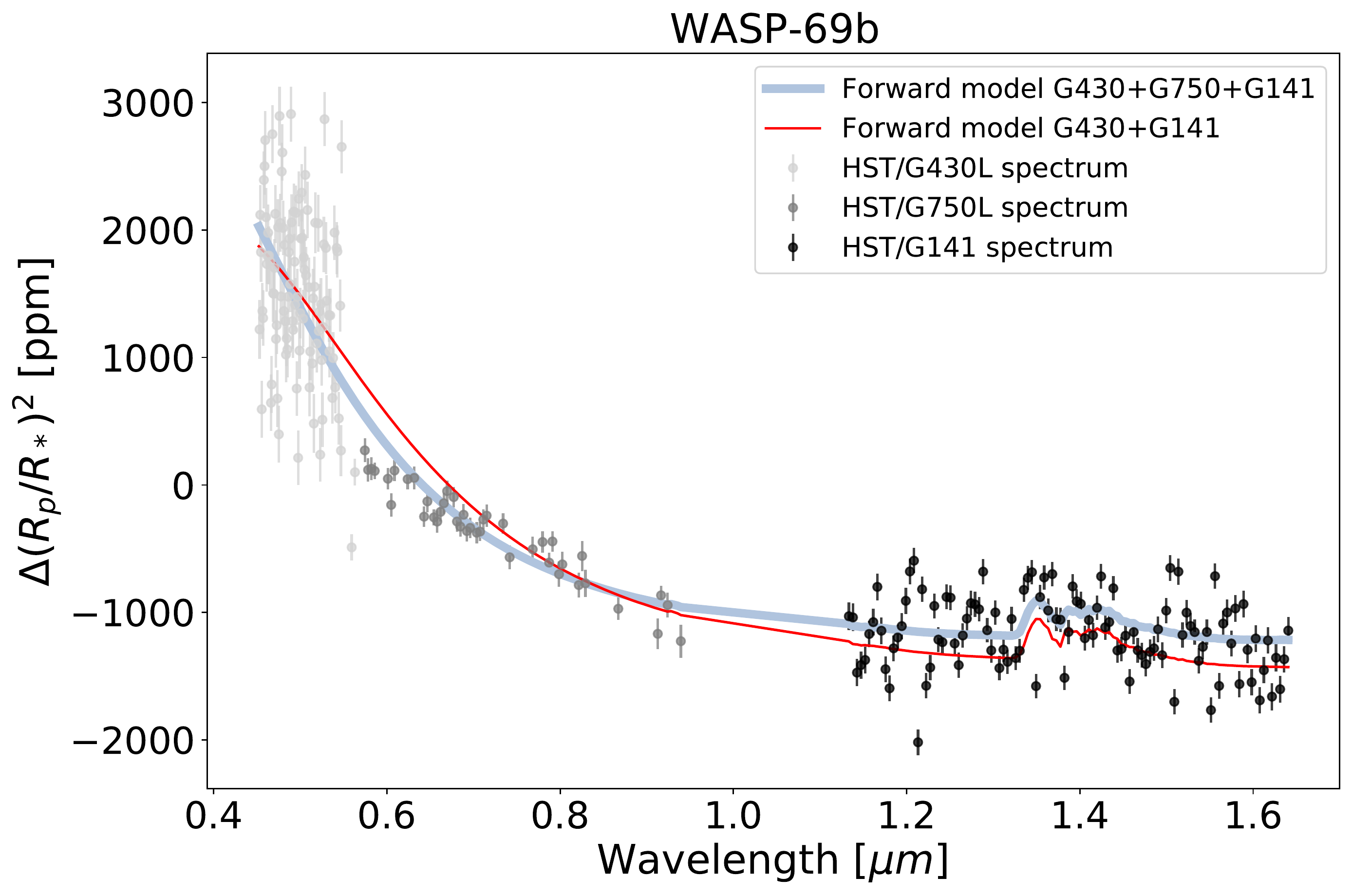}
\caption{Comparison between the forward models based on the maximum likelihood of the retrieval obtained with CERBERUS for HST/G430L+G141 (red) and the full spectra (G430L+G750L+G141L, in blue).}
\label{fig:comparison}
\end{figure}

Our observations in HST/G750L are not consistent with previous ground-based measurements in the same wavelength band. We performed a Kolmogorov–Smirnov (KS) test between our HST/G750L data and the ground-based data and found that the null hypothesis that the two distributions are similar can be rejected as the p-value found by the test is lower than the level of significance (a = 0.05). Divergence between space- and ground-based observations in the optical have been reported before in the literature for the exoplanet WASP-121b \citep{Wilson21}.
There are at least three factors that can explain the inconsistency between the slopes. 
\begin{enumerate}
    \item The divergence at short wavelengths can be due to the treatment of the limb darkening, which can introduce small discrepancies, affecting mostly short wavelengths. Our limb-darkening correction is based on the four parameter non-linear limb-darkening model which is described by \cite{Claret00} to be a more accurately approach to estimate the intensity distribution than the quadratic law adopted by the ground-based work of \cite{Murgas2020}. 
    \item Inconsistencies in the retrieved transit depth could be due to differences in the orbital system parameters assumed. Here, we used the system parameters from \cite{Stassun17}, with the exception of the semi-major axis that we kept fixed to the value from the discovery paper \citep{And14}. We keep the orbital period, eccentricity, and semi-major axis fixed to the literature, and fit for the planet-to-star radius ratio (R$_{p}$/R$_{*}$), time of the mid transit (T$_{mid}$) and inclination ($i$), as shown in Table \ref{tab:system}. By contrast, \cite{Murgas2020} fixed the period and orbital eccentricity to the values from \cite{And14}. 
    \item Finally, the strength of the analysis presented here is given by the consistency between HST/G430L+G141 and the full merged spectra (G430L+G750L+G141) retrieval; the high degree of similarity in the aerosol model solution with and without the G750L data demonstrates that the G430L and G750L data are in agreement in terms of the underlying physical effect they measure (see Fig. ~\ref{fig:comparison}). Said differently, the HST/G750L+G141 spectrum, which has the same wavelengths as \cite{Murgas2020}, is not driving the retrieval results of the full spectrum. Also, our analysis has the absence of uncertainty introduced by the telluric correction, which is a factor that could influence ground-based observations.  \\
\end{enumerate}

\begin{table}[!ht]
\caption{Orbital system parameters of WASP-69b retrieved with CERBERUS and limb-darkening coefficients.}
\label{tab:system}
\centering
\begin{tabular}{ll}
Transit Parameter                                      & Value                          
\\
\hline
R$_{p}$/R$_{*}$                                        & 0.133 $\pm$ 0.00004                                           \\
a$_{p}$/R$_{*}$                                       & 0.04525 $\pm$ 0.00053 (fixed)$^{a}$                           \\
T$_{mid}$ {[}BJD$_{TBD}${]}                            & 55748.32 $\pm$ 0.00018                                        \\
Period [days]                                         & 3.86814 $\pm$ 2$\times 10^{-6}$ (fixed)$^{b}$                             \\
i (deg)                                                & 86.10 $\pm$ 0.0029                                            \\
e                                                      & 0 $\pm$ 0$^{b}$ (fixed)                                       \\
u$_{1}$                                                & 2.19                                                          \\
u$_{2}$                                                & -3.85                                                         \\
u$_{3}$                                                & 4.08                                                          \\
u$_{4}$                                                & -1.41                                                         \\
\hline
\multicolumn{2}{l}{\begin{tabular}[c]{@{}l@{}}$^{a}$\cite{And14}\\ $^{b}$ \cite{Stassun17}
\end{tabular}}
\end{tabular}
\end{table}



\subsection{Merged Spectral Observations}
The transmission spectra show that the planet/star ratios measured with HST/STIS is larger than that obtained with HST/WFC3; a key question to address in our analysis is whether these differences reflect intrinsic properties of the atmosphere or are differences due to different instrument systematics or stellar activity. Although previous studies \citep{Sing16, Chachan19, Rath21} show broad spectral coverage by combining HST and Spitzer measurements, they did not take this aspect into account in their atmospheric retrieval. Here, we use CERBERUS to retrieve the offsets between the filters, using the G430L filter as a reference to shift STIS/G750L (offset0) and WFC3/G141 (offset1) accordingly. The retrieval finds a well-constrained distribution for the offsets, indicating that the offsets between G430L\textendash G750L and G430L\textendash G141 are small, -0.006\% (-60 ppm)  and -0.01\% (-100 ppm), respectively. In their previous work on WASP-69b, \cite{Murgas2020} found an offset of -470 ppm between their ground-based optical observations and the NIR data from HST/WFC3. Our retrieval found a -40 ppm offset for the same spectral region (G750L\textendash G141). The incompatibility between the offsets could be related to two factors: i) \cite{Murgas2020} performed the retrieval of the offsets in a joint spectra between two different instruments, Gran Telescopio Canarias (GTC)/OSIRIS and HST, where the individual data sets were reduced separately. The HST and GTC data used different limb darkening parameters (\cite{Tsiaras18} and \cite{Murgas2020}, respectively) and they may not have had self-consistent system parameters.; ii) there are potentially data processing differences between the GTC/OSIRIS, reduced by \cite{Murgas2020}, and HST data, reduced by \cite{Tsiaras18}, that could lead to discrepancies between their results. In contrast, our HST data reduction is performed in a uniform and self-consistent way for all filters (G430L, G750L and G141).\\

\begin{table*}[]
\caption{Atmospheric model constraints estimated from the transmission spectrum by spectral retrieval. The optical depth scale factor is a multiplicative factor of Jupiter's haze layer optical depth ($\tau_{\rm Jup}$). }
\label{tab:retrieval}
\begin{tabular}{lllll}
Parameter                          & Label  & Median value                  & 16th percentile        & 84th percentile    \\
\hline
Cloud top pressure                 & CTP    & 0.04bar                & 0.03bar                & 0.05bar            \\
Haze optical depth scale factor    & HScale & 10$^{-3.49}$ $\tau_{Jup}$     & 10$^{-3.54}$ $\tau_{Jup}$    & 10$^{-3.40}$ $\tau_{Jup}$ \\
Haze profile peak density pressure & HLoc   & $<$ 5.3$\mu$bar            & --           & --      \\
Haze vertical extent scale factor  & HThick & 10$^{10.91}$                 & 10$^{7.30}$              & 10$^{16.76}$          \\
isothermal temperature profile     & T      & $>$ 1450K                  & --                 & --              \\
Metallicity [X/H]                  & TEC[0] & 10$^{-2.79}$              & 10$^{-2.92}$             & 10$^{-2.36}$              \\
C/O                                & TEC[1] & 10$^{-1.51}$            & 10$^{-3.25}$          &   10$^{0.23}$             \\
N/O                                & TEC[2] & 10$^{-1.56}$              & 10$^{-4.15}$       &   10$^{1.19}$             \\
Offset0                            & Off0   & -0.006\%                & -0.007\%                & -0.004\%            \\
Offset1                            & Off1   & -0.010\%                & -0.011\%                & -0.008\%           
\end{tabular}
\end{table*}






\subsection{Stellar Activity Impact}
Presence of unocculted spots and/or faculae on the visible stellar disk has been shown to introduce wavelength dependent variability in the measured relative planetary radius \citep{Rack18}, that could be interpreted as a signal of atmospheric origin. As WASP-69 is a relatively active star showing strong emission in the Ca II H+K line cores ($\log$\,$R^\prime_{HK}$\,$\sim$\,$-4.54$), with photometric variability of $\sim$20\,mmag \citep{And14}, such effects need to be considered when interpreting its transmission spectrum. In order to estimate the impact of heterogeneities on the stellar surface upon the observed transmission spectrum, we perform a retrieval analysis of the combined spectrum where the planet is assumed to contain no optically thin atmosphere. Namely, its radius is assumed to be constant at all wavelengths and active regions on the star are entirely responsible for the morphology observed in the spectrum. The retrieval has been performed using the PLATON package \citep{Zhang19}, where the contamination spectrum is modelled as:
\begin{equation}
    \left(\frac{R_p}{R_\star}\right)^2_{\lambda,c} =  \left(\frac{R_p}{R_\star}\right)^2_{\lambda} \frac{S(\lambda,T_{het})}{f_{het} S(\lambda,T_{het}) + (1-f_{het})S(\lambda,T_\star)}
\end{equation}
\noindent For the retrieval, the constant planetary radius ($R_p$), fractional heterogeneity coverage ($f_{het}$) and the active region temperature ($T_{het}$) are adopted as free parameters with uninformative prior distributions assumed. Furthermore, we also adopted the photospheric temperature ($T_\star$) as a free parameter, but under the assumption of a Gaussian prior \citep[$\mathcal{N}(4715\,K, 200)$;][]{And14}. In this framework $S(\lambda,T)$ are the interpolated BT-NextGen (AGSS2009) stellar models of \cite{Allard12}. The posterior space is sampled using a nested sampling algorithm with 2000 live points.

\begin{figure}[!ht]
\centering
\includegraphics[width=0.5\textwidth]{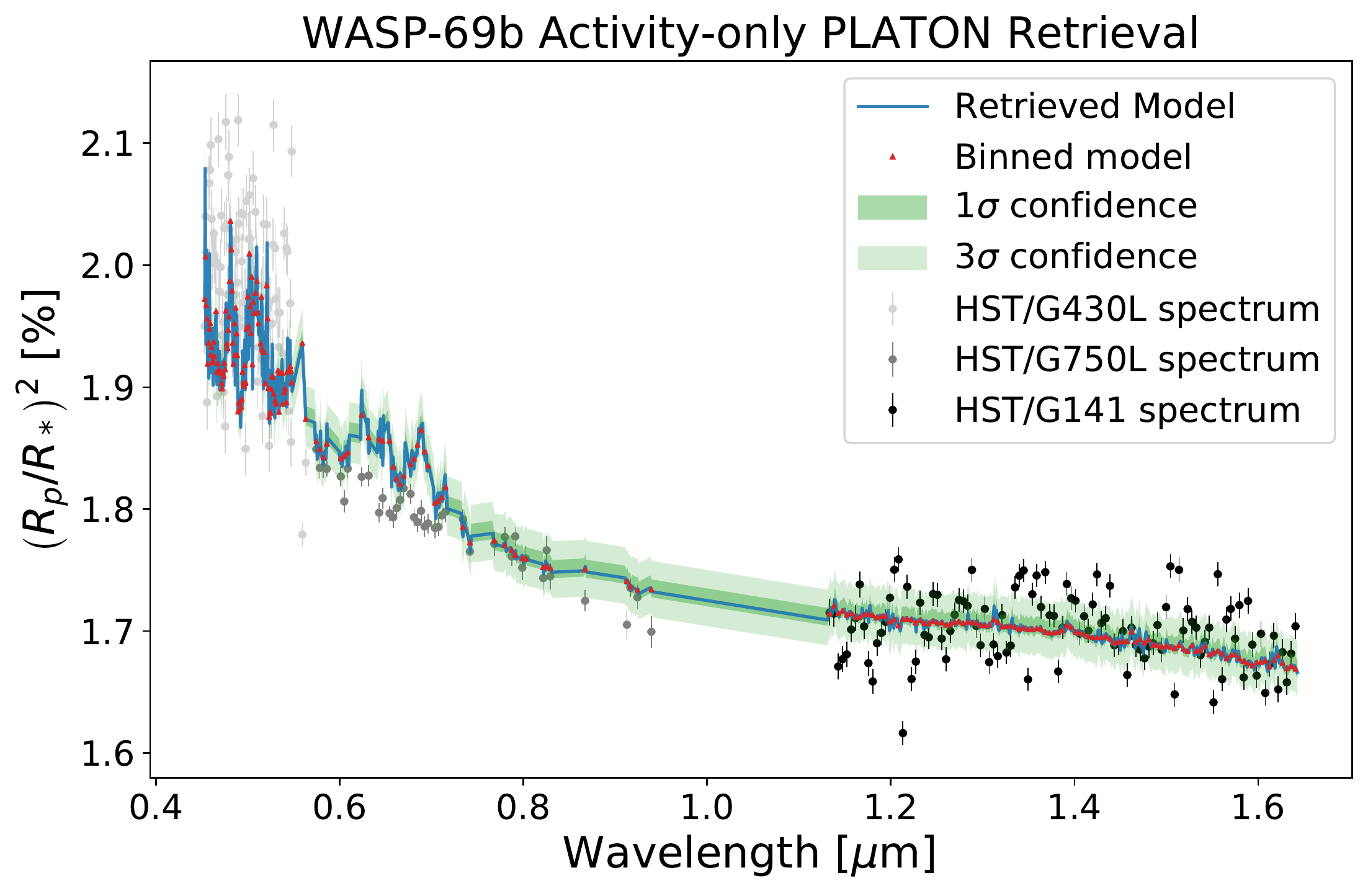} 
\caption{PLATON activity-only retrieved model for the composite HST transmission spectrum of WASP-69b, shown as the solid blue line, with the 1- and 3-$\sigma$ confidence intervals highlighted in green. The binned model within the spectrophotometric channels are presented as red triangles.}
\label{fig:activitymodel}
\end{figure}

\begin{figure}[!ht]
\centering
\includegraphics[width=0.5\textwidth]{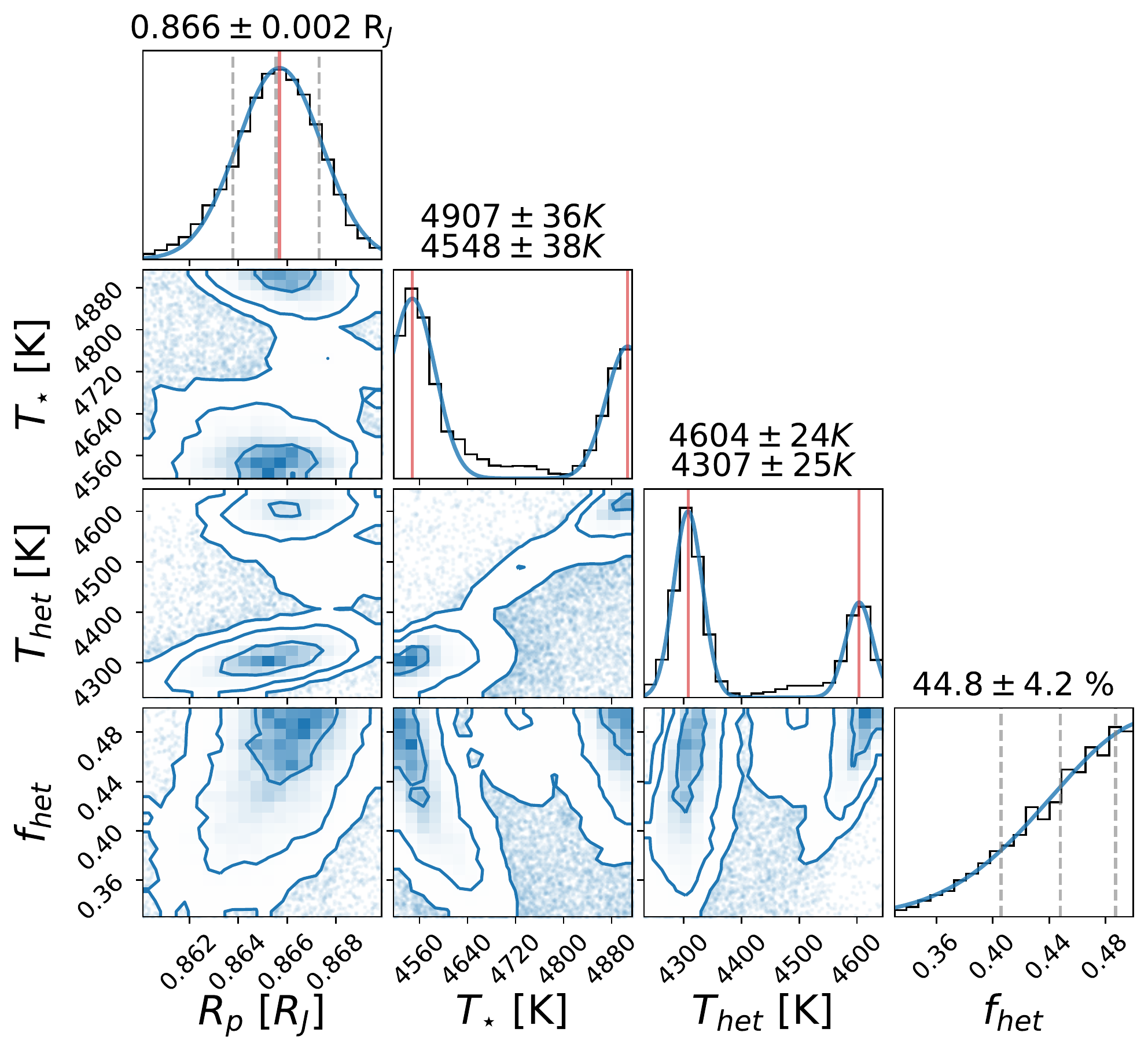} 
\caption{Correlations and marginalised distributions from posterior samples obtained with nested sampling run of 2000 live points, for the activity-only retrieval of HST spectrum, which was presented in Fig. \ref{fig:activitymodel}. Vertical red lines in the marginalised plots show the modes of the fitted Gaussian distributions (solid blue lines), with the gray dashed lines highlighting the 1-$\sigma$ confidence intervals. These have been omitted from the temperature plots for clarity. The unconstrained distribution of the fractional coverage, $f_{het}$ means that the mode lies beyond the prior limits and therefore the median is presented instead.}
\label{fig:activity posterior}
\end{figure}

This best fit model is shown in Fig. \ref{fig:activitymodel}, where the slope towards the near-UV is retrieved relatively well by this activity model. Posterior samples and their corresponding marginalised distributions for the adopted free parameters are shown in Fig. \ref{fig:activity posterior}, from which best fit solutions are derived. We obtain a constant planetary radius of 0.866\,$\pm$\,0.002\,R$_J$ for this activity-only model, with an unconstrained posterior distribution for the fractional coverage, $f_{het}$, whose modal value (49.4\,$\pm$\,4.1\,\%) reaches the limit of what is physically accepted. The median of this factor (44.8\,$\pm$\,4.2\,\%) still points to very large coverage of the stellar disk by cooler active spots, which is marginally beyond the limit of what is expected for cooler dwarfs \citep{Jackson13}. We observe bimodal distributions for the photospheric and active-region temperatures owing to inherent degeneracies, which has been observed in other stars \citep[e.g.][]{Espinoza19}. The most probable solution is a stellar photospheric temperature of 4550\,$\pm$\,40\,K with spots temperature of 241\,$\pm$\,25\,K below the temperature of the surrounding photosphere. The other possible, but less likely, solutions for the photosphere and spot temperatures are also annotated in Fig. \ref{fig:activity posterior}. Such large spot coverage, owing either to the presence of many solar-type or giant spots, was also corroborated by \cite{Murgas2020}.


However, we find no evidence for star spot crossing in our three HST transit light curves. Similarly, observations by \cite{Murgas2020} using GTC/OSIRIS also found no evidence for star spot crossing in previous transits. Regardless, \cite{Murgas2020} do not discard the possibility of spots because their data have a short interval of only 50 days. In our data, it is unlikely that the same group of spots would be causing the slope in both STIS and WFC3, as our STIS/G750L data taken on 4 October 2017, is $\sim$395 days apart from WFC3/G141 (16 August 2016). This interval is longer than active region lifetime of 47 $\pm$ 5 days found for the active G-type star Kepler-17 \citep{Bonomo12}. While observed lifetime of spots on the K-type star Kepler-210 could be even lower, of about 60\textendash90 days \citep{Ioannidis16}, which is expected as starspots decay more slowly on cooler stars and the lifetime of the spots for K dwarfs can vary between 50 and 100 days \citep{Giles17}.

There are a total of 4 transit light curves including the GTC observations, none of which show evidence for star spot crossings, which argues against high levels of spot coverage. A delta $\chi^{2}$ test shows that the CERBERUS model is strongly preferred to the stellar spot model. This, in combination with the absence of evidence of spots in the transit light curve, suggests that stellar activity is not responsible for the measured transit spectrum. Thus, similar to \cite{Murgas2020}, we conclude that stellar activity is less likely to be responsible for the observed transmission spectrum which is more likely to be consistent with Rayleigh scattering due to hazes.


\section{Discussion}
\label{sec:disc}


Our observational detection of aerosol-induced modulation in the spectra of WASP-69b highlights three important aspects of hazes in exoplanets atmospheres: 
\begin{enumerate}
    \item Haze can exist at microbar pressures and that confirms a number of models that have suggested the presence of haze at high altitudes \citep{Lavvas17,Lavvas19,Adams19,Gao2020},
    \item The haze layer is approximately uniformly distributed throughout the atmospheric column. Constant haze density profiles of a wide range of pressures have been described in the literature \citep{Lavvas17} and is associated with a high temperature profile and a larger atmospheric mixing,
    \item The straight line in the optical slope that we see in previous atmospheric models \citep{Mc14,Wakeford15,Sing16,Murgas2020} is not the only possibility. Previous work \citep{Pinhas17} has discussed the possible association of a range of condensates for producing a non-linear, or curved, Rayleigh slope.
    Our atmospheric retrieval CERBERUS models the Rayleigh slope with a curved shape slope, in which the shape of the curve is controlled by the multiplicative factor of Jupiter's optical depth (HScale) and the peak density of the aerosol profile. When the haze is a distinct layer, by simulating a model with a slightly increased HScale and HLoc set up to lower pressures, the aerosol Rayleigh scattering slope is closer to a straight line. 
\end{enumerate}

Aerosols at high altitudes have been suggested as a possible explanation for planets showing a prominent scattering slope in the optical and NIR of their transmission spectra. Particularly another warm sub-Saturn, HAT-P-12b \citep{Wong20}, has a similar radius, mass, and orbital period (M$_{p}$ = 67.05 M$_{\oplus}$ , R$_{p}$ = 10.79 R$_{\oplus}$, and P = 3.21 days) to WASP-69b and also orbits a K dwarf \citep{Hart09}. Analysis of the size and distribution of the haze for this target indicates that haze would be formed by sub-micron particles, with the higher densities peaked towards lower pressures, and the highest pressure probed by their observations at $\sim$10$^{-4}$ bars. However, our work provides direct evidence of aerosols at microbars, and a greater sampling of an exoplanet's atmosphere.

It has also been suggested that hydrogen/helium atmospheric escape can possibly produce slopes towards the visible \cite{Murgas2020}. WASP-69b was initially thought to be losing its atmosphere in a comet-like form that extends behind the planet after detection using the helium triplet at 1083 nm \citep{Nort18}. More recent observations \citep{Vissapragada2020} using better sampled light curves and higher precision instruments found no evidences of escape. Simulations during planetary transit of the ionized mass loss and the helium absorption of WASP-69b are consistent with the symmetric transit shape and do not show any prominent comet-like tail \citep{Wang21}. These studies conclude that WASP-69b is likely not undergoing a strong escape and therefore the slope towards the visible in its transmission spectra is not likely associated with a strong escape.

\begin{figure}[!ht]
\centering
\includegraphics[width=0.5\textwidth]{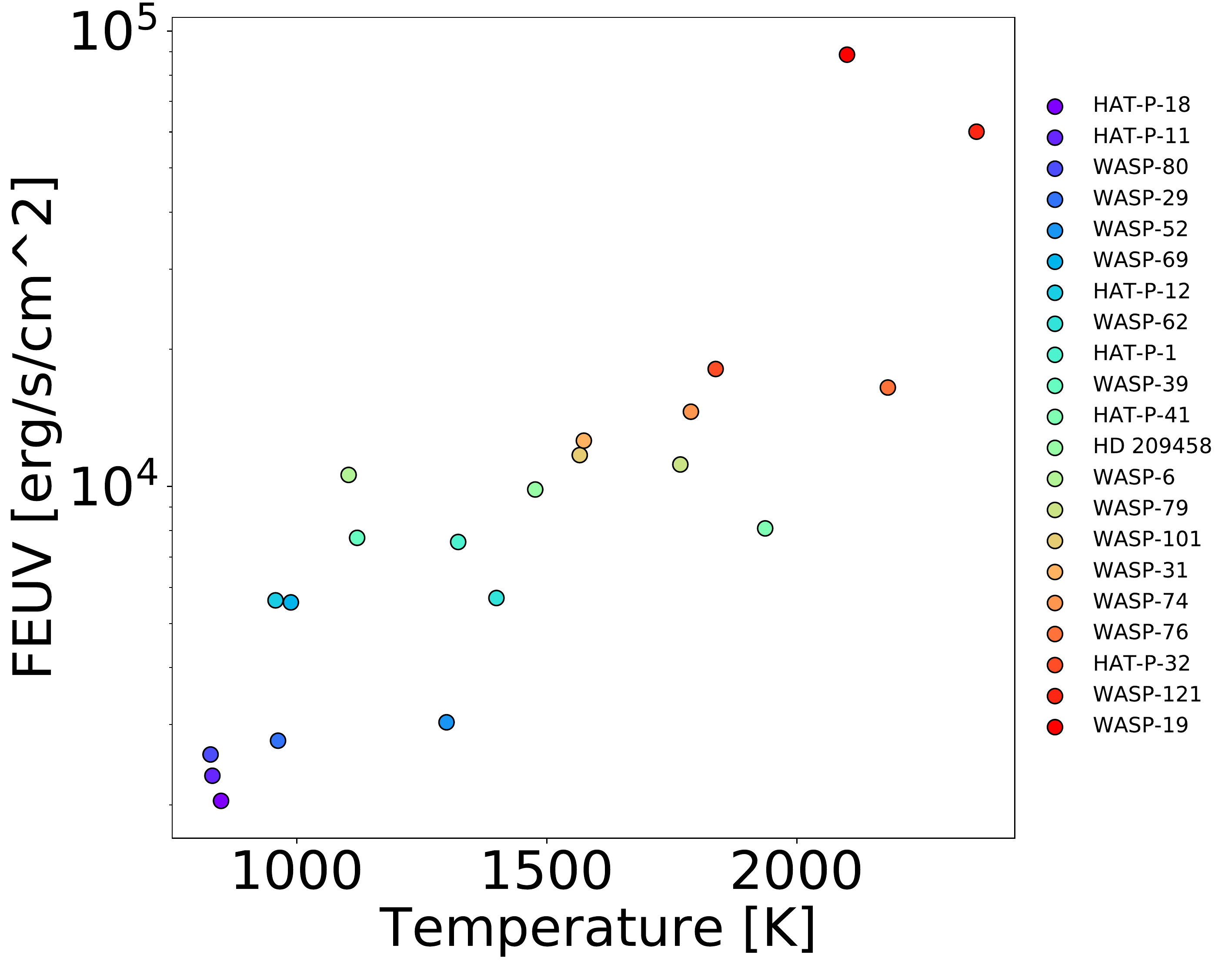}
\caption{Extreme ultraviolet fluxes as a function of temperature for the giant planets observed with HST/STIS.}
\label{fig:FEUVfluxes}
\end{figure}

Aerosol composition is still an open debate and not constrained by the data we analyzed. In principle, the scattering could be due to photochemical haze produced from hydrocarbon aerosols or it could be due to condensates. Although our work does not directly answer this question, we investigated some factors, such as extreme-ultraviolet flux (FEUV), C/O and temperature, that could help in determining aerosol composition. We estimated that WASP-69b receives an FEUV of 5563 erg/s/cm$^{2}$ \citep{Estrela20}, which is approximately 400 times stronger than the flux received by Earth. This FEUV level is moderate compared to other hot Jupiters observed with HST/STIS, as some of these other hot Jupiters receive even stronger FEUV fluxes, as shown in Fig. \ref{fig:FEUVfluxes}. Planets with high irradiation levels of FEUV, such as WASP-19b, show evidence of a slope towards the blue \citep{Sing16,Seda17,Seda21} but covering less scale heights than the spectra in WASP-69b. These observations lead to another question: if photochemical haze produces the Rayleigh scattering observed in WASP-69b, why is it probing more scale heights in this planet than in other observations of more strongly irradiated planets? Also, laboratory studies of haze formation in hot Jupiter atmospheres show that photochemical haze formation occurs when the C/O $>$ 1 and is less efficient for C/O $<$ 1 \citep{Fleury19, Fleury20}. Although the C/O retrieved here for WASP-69b does not have tight constraints, there is some preference for C/O lower than  sub-solar values. These two points suggest that another mechanism besides photochemistry could be inducing the haze formation in WASP-69b. However, further work is required to understand aerosol composition in this atmosphere.

Previous studies have already suggested that hydrocarbons in close-in giant planets are significantly lower than the quantities present on Jupiter and Saturn \citep{Liang04}, in which the formation of hydrocarbon aerosols is the dominant source of photochemical hazes. It is also possible that the scattering slope is due to condensates \citep{Lecavelier08}. Multiples condensates (Si-O, Al-O, Fe-O, Ti-O) have been proposed for hot Jupiters with strong scattering, such as HD 189733b which shows strong scattering in the optical to the NIR due to Rayleigh scattering covering 7 H$\rm _{s}$ of spectral modulation \citep{Pont13,Mc14}. 
In particular, silicates (MgSiO$_{3}$ and MgSiO$_{4}$, Fe poor) show strong Rayleigh scattering slopes at short wavelengths up to 3 $\rm \mu$m and could be potential candidates to explain the slope in WASP-69b \citep{Wakeford15, Pinhas17}. These silicates are believed to be the main opacity source for giant planets with equilibrium temperature (Teq) above 950 K. The condensation curves of silicates MgSiO$_{3}$ and MgSiO$_{4}$ indicate that these aerosols condense at temperatures around 1400\textendash1500 K at lower pressures \citep{Morley12}. We retrieved atmospheric temperatures $>$ 1460 K which, as mentioned previously, are consistent with silicate condensation. These temperatures are higher than the Teq of the planet and could be also related to photochemical haze warming the upper atmosphere as proposed recently by \cite{Lavvas21}.

\section{Conclusions}
We present here the detection of a highly extended atmosphere in the warm sub-Saturn WASP-69b using a combined effort of HST/STIS and HST/WFC3 observations. The spectral modulation covers 11 H$\rm _{s}$ representing a greater sampling of the atmosphere of an exoplanet than any previous measurement. This modulation is induced by an approximately uniform haze density that extends throughout the sampled atmospheric column from 40 milibar to microbar pressures. Our results are an observational confirmation of previous microphysics models in the literature that suggested aerosols at microbar pressures. Although our work does not specifically enable analysis of this aerosol-dominated atmosphere, we believe it will be useful in producing STIS spectra at relatively high resolution in future studies of planets with strong K or Na features.


The aerosol composition of Wasp-69b is still unresolved, but two possible sources are photochemical haze produced from hydrocarbon aerosols or condensates. Our tentative C/O ratio suggests sub-solar values, which is indicative of condensates such as MgSOx, already identified as a possibility in other studies as a dominant opacity in hot giant exoplanets with equilibrium temperatures above 950 K \citep{Wakeford15, Gao2020}. However without significant vertical transport, it seems unlikely that a condensate can be present at such low pressures. Although the question of the origin of this aerosol cannot be decisively answered by our observations, this work identifies this exoplanet as an important object for follow up studies. James Webb Space Telescope (JWST) observations, with broader spectral coverage, could establish the C/O ratio and hence the possible origin (and potentially atmospheric mixing values) of aerosols on this exoplanet. 

\section*{Acknowledgments}

This research has made use of the NASA Exoplanet Archive, which is operated by the California Institute of Technology, under contract with the National Aeronautics and Space Administration under the Exoplanet Exploration Program. Raissa Estrela, Mark Swain acknowledge support for a portion of this effort from NASA ADAP award 907524. Raissa Estrela acknowledges Sao Paulo Research Foundation (FAPESP) for the fellowship \#2018/09984-7. This work has been supported in part by the California Institute of Technology Jet Propulsion Laboratory Exoplanet Science Initiative. This research was carried out at the Jet Propulsion Laboratory, California Institute of Technology, under a contract with the National Aeronautics and Space Administration (80NM0018D0004). © 2021.
All rights reserved. 

\software{CERBERUS \citep{Roudier21}, PyMC3 \citep{Salvatier16}, PLATON \citep{Zhang19}}

\bibliography{sample631}{}
\bibliographystyle{aasjournal}



\end{document}